\documentclass[
aps,
prd,
showpacs,
preprint,
tightenlines,
superscriptaddress,
nofootinbib,
floatfix,
a4paper]
{revtex4}
\usepackage{graphicx}
\begin{document}
\title{  Addendum to:\\ Model-dependent and -independent
implications\\ of the first Sudbury Neutrino Observatory results}
\author{        G.L.~Fogli}
\affiliation{   Dipartimento di Fisica
                and Sezione INFN di Bari\\
                Via Amendola 173, 70126 Bari, Italy\\}
\author{        E.~Lisi}
\affiliation{   Dipartimento di Fisica
                and Sezione INFN di Bari\\
                Via Amendola 173, 70126 Bari, Italy\\}
\author{        D.~Montanino}
\affiliation{   Dipartimento di Scienza dei Materiali
                and Sezione INFN di Lecce\\
                Via Arnesano, 73100 Lecce, Italy\\}
\author{        A.~Palazzo}
\affiliation{   Dipartimento di Fisica
                and Sezione INFN di Bari\\
                Via Amendola 173, 70126 Bari, Italy\\}

\begin{abstract}
In the light of recent experimental and theoretical improvements,
we review our previous model-independent comparison \cite{Ours} of
the Super-Kamiokande (SK) and Sudbury Neutrino Observatory (SNO)
solar neutrino event rates, including updated values for the
``equalized'' SK datum and for the reference Standard Solar Model
(SSM) $^8$B neutrino flux. We find that the joint SK~+~SNO
evidence for active neutrino flavor transitions is confirmed at
the level of $3.3$ standard deviations, independently of possible
transitions to sterile states. Barring sterile neutrinos, we
estimate the $\pm 3\sigma$ range for the $^8$B neutrino flux
(normalized to SSM) as $f_B=0.96^{+0.54}_{-0.55}$. Accordingly,
the $\pm 3\sigma$ range for the energy-averaged $\nu_e$ survival
probability is found to be $\langle P_{ee}\rangle =
0.31^{+0.55}_{-0.16}\,$, independently of the functional form of
$P_{ee}$.  An increase of the reference $\nu_e+ d\to p +p+ e^-$
cross section by $\sim 3\%$, as suggested by recent theoretical
calculations, would slightly shift the central values of  $f_B$
and of $\langle P_{ee}\rangle$ to $\sim 1.00$ and $\sim 0.29$,
respectively, and would strengthen the model-independent evidence
for $\nu_e$ transitions into active states at the level of $\sim
3.6\, \sigma$.
\end{abstract}
\medskip
\maketitle

\section{Introduction}

In a previous paper \cite{Ours} we have used the ``threshold
equalization'' technique introduced by Villante {\em et al.\/} in
\cite{Vill} to perform a model-independent comparison of the solar
neutrino event rates measured in Super-Kamiokande (SK) \cite{SKol}
through $\nu$-$e$ elastic scattering (ES), and in the Sudbury
Neutrino Observatory (SNO) \cite{SNO1} through charged-current
(CC) $\nu_e$ absorption in deuterium. In particular, it was shown
in \cite{Ours} that the energy spectrum of parent neutrinos in SNO
(namely, the spectrum of neutrinos producing an electron with {\em
measured\/} kinetic energy $T_e^\mathrm {SNO} \geq 6.75$ MeV) is
accurately reproduced by the analogous spectrum in SK, provided
that the corresponding SK threshold is shifted from its default
value ($T_e^\mathrm{SK} \geq 5 \mathrm{\ MeV}-m_e$) to the
``equalization value'' $T_e^\mathrm{SK} \geq  8.6$ MeV (see also
\cite{Vill,Equa,Fior}).

On this basis, we have quantified in \cite{Ours} the joint SK+SNO
evidence for $\nu_e\to\nu_{\mu,\tau}$ transitions at the level of
$3.1\sigma$, {\em independently\/} \cite{Vill} of (i) the Standard
Solar Model (SSM) prediction $\Phi_B^\mathrm{SSM}$ for the $^8$B
neutrino flux; (ii) the functional form of the $\nu_e$ survival
probability $P_{ee}$; and (iii) the presence of possible
transitions \cite{Ster,Pena} to a sterile state $\nu_s$. Barring
$\nu_e\to\nu_s$ transitions, we also estimated in \cite{Ours} the
allowed ranges for the {\em true\/} $^8$B neutrino flux $\Phi_B$
(or, equivalently, for $f_B=\Phi_B/\Phi_B^\mathrm{SSM}$), and for
$\langle P_{ee}\rangle$ (averaged over the equalized neutrino
energy spectrum). The results in \cite{Ours} indicated a strong
$\nu_e$ suppression at the Earth ($\langle P_{ee}\rangle \sim
0.3$) and a striking agreement of $\Phi_B$ with the reference SSM
prediction of \cite{BP00}.

After the work \cite{Ours}, there have been several improvements
(at the level of $\lesssim 1\sigma$) in the quantities of interest
for a model-independent analysis, including (a) an updated value
for the SK flux $\Phi_\mathrm{ES}^\mathrm{SK}$ above the usual
threshold \cite{Noon,Kosh}, as well as a preliminary evaluation of
the corresponding value above $T_e^\mathrm{SK} \geq 8.6$ MeV
\cite{Priv}; (b) an updated reference value for
$\Phi_B^\mathrm{SSM}$ \cite{Robu} following the recent, accurate
evaluation of the astrophysical $S_{17}(0)$ factor in \cite{S170};
and (c) theoretical corrections \cite{Beac,Voge,Naka} to the
reference $\nu_e$-$d$ CC cross section \cite{Xsec}.

Since the significance of the SNO-SK model-independent results in
\cite{Ours} was just above $3\sigma$, we think it relevant to
check whether the above $\lesssim 1\sigma$ variations in input
quantities conspire (or not) to weaken such significance.%
\footnote{Evidence for new physics at the level of $\sim 3\sigma$
may be fragile under small changes or corrections to input
quantities. A recent example is provided by the reduction of the
muon $g-2$ anomaly \cite{Muon}.}
After a brief review of the previously adopted reference neutrino
fluxes (Sec.~II), we include the most recent improvements in the
model-independent analysis (Sec.~III), and show that the results
in \cite{Ours} are actually strengthened. We draw our conclusions
in Sec.~IV.
\footnote{In this {\em addendum\/} to \cite{Ours}, we do not
discuss how model-dependent results (i.e., oscillation analyses
\cite{Ours,Pena,Fits}) are modified by (some of) the updated
inputs; for such issues see, e.g., the recent works
\cite{Noon,Robu,Conc}.}

\section{Previous reference values and results}

In the previous model-independent analysis \cite{Ours}, we
considered the following reference neutrino fluxes:
\begin{eqnarray}
\Phi_\mathrm{CC}^\mathrm{SNO}   &=& 1.75 \pm 0.148\
(T_e^\mathrm{SNO}\geq 6.75\mathrm{\ MeV})\ ,             \label{e1}\\
\Phi_\mathrm{ES}^\mathrm{SK}    &=& 2.32 \pm 0.085\
(T_e^\mathrm{SK}\geq 5\mathrm{\ MeV}-m_e)\ ,             \label{e2}\\
\Phi_\mathrm{ES}^\mathrm{SK}    &=& 2.28 \pm 0.085\
(T_e^\mathrm{SK}\geq 8.6\mathrm{\ MeV})\ ,                \label{e3}\\
\Phi_{B}^\mathrm{SSM}           &=& 5.05\times
 (1^{+0.20}_{-0.16})\ ,                                  \label{e4}
\end{eqnarray}
expressed in units of $10^6$~cm$^{-2}$~s$^{-1}$, and with attached
$\pm 1\sigma$ uncertainties.

In the above equations, $\Phi_\mathrm{CC}^\mathrm{SNO}$ represents
the measured flux of $\nu_e$ inducing CC events in SNO
\cite{SNO1}, whereas $\Phi_\mathrm{ES}^\mathrm{SK}$ is the flux of
active $\nu$'s inducing ES events in SK. In particular, the value
of $\Phi_\mathrm{ES}^\mathrm{SK}$ in Eq.~(\ref{e2}) refers to a
threshold of 5 MeV for the electron total energy, and to 1258 days
of SK detector exposure \cite{SKol}, while the value in
Eq.~(\ref{e3}) refers to a threshold of 8.6 MeV for the electron
kinetic energy (so as to equalize the SK and SNO response
functions \cite{Ours,Vill,Equa}), as derived from a SNO reanalysis
of the SK
spectrum data \cite{Ours,SNO1}.%
\footnote{The numbers in Eq.~(\ref{e3}) have neither been disputed
nor (unfortunately) confirmed or re-estimated by the SK
collaboration, as far as the 1258 day data sample is concerned.}
Finally, the value of $\Phi_B^\mathrm{SSM}$ in Eq.~(\ref{e4})
represents the reference $^8$B $\nu$ flux from the so-called
``BP00'' SSM \cite{BP00}.

The model-independent analysis in \cite{Ours} was based on the
central values and errors in Eqs.~(\ref{e1}) and (\ref{e3}),
which, following common practice, were conventionally normalized
to the {\em central value\/} in Eq.~(\ref{e4}), giving
\begin{eqnarray}
\Phi_\mathrm{CC}^\mathrm{SNO}/\Phi^\mathrm{SSM}_B
&=& 0.347 \pm 0.029 \label{snold}\ ,\\
\Phi_\mathrm{ES}^\mathrm{SK}/\Phi^\mathrm{SSM}_B &=& 0.451 \pm
0.017 \label{skold}\ ,
\end{eqnarray}
whose $3.1 \sigma$ relative difference provided a completely
model-independent evidence of $\nu_e$ transitions to active
states, even in the presence of possible sterile neutrino
$(\nu_s)$ mixing. Such evidence was also illustrated in Fig.~2
of \cite{Ours}.%
\footnote{In \cite{Ours}, the flux ratios in Eqs.~(\ref{snold})
and (\ref{skold}) were also indicated as SK/SSM and SNO/SSM,
respectively.}

Assuming only active transitions, the above flux ratios
(corresponding to equalized SNO and SK spectra) are linked by the
following {\em exact\/} relations \cite{Ours,Vill,Equa})
\begin{eqnarray}
\Phi_\mathrm{CC}^\mathrm{SNO}/\Phi^\mathrm{SSM}_B
&=& f_B \, \langle P_{ee}\rangle \ ,\label{snold1}\\
\Phi_\mathrm{ES}^\mathrm{SK}/\Phi^\mathrm{SSM}_B &=& f_B\,
(0.848\, \langle P_{ee}\rangle+0.152)
 \label{skold1}\ .
\end{eqnarray}
These relations, together with Eqs.~(\ref{snold}) and
(\ref{skold}), allowed to derive in \cite{Ours} the $\pm 3\sigma$
range for the {\em true\/} $^8$B neutrino flux $\Phi_B$
[normalized to the central value in Eq.~(\ref{e4}),
$f_B=\Phi_B/\Phi_B^\mathrm{SSM}$],
\begin{eqnarray}
f_B  &=& 1.03^{+0.50}_{-0.58} ,\label{fBold}
\end{eqnarray}
as well as the analogous range for the $\nu_e$ survival
probability $\langle P_{ee}\rangle$, averaged over the (equalized
\cite{Vill,Equa}) parent neutrino energy spectrum,
\begin{eqnarray}
\langle P_{ee}\rangle  &=& 0.34^{+0.61}_{-0.18} .\label{Peeold}
\end{eqnarray}
The BP00 standard solar model  and the standard electroweak model
appeared then to be respectively confirmed and disconfirmed by the
above results, as also graphically shown in Fig.~3 of \cite{Ours}.

\section{Updated reference values and results}

Let us consider now the following reference neutrino fluxes (in
units of $10^6$~cm$^{-2}$~s$^{-1}$),
\begin{eqnarray}
\Phi_\mathrm{CC}^\mathrm{SNO}   &=& 1.75 \pm 0.148\
(T_e^\mathrm{SNO}\geq 6.75\mathrm{\ MeV})\ ,             \label{e1new}\\
\Phi_\mathrm{ES}^\mathrm{SK}    &=& 2.35 \pm 0.085\
(T_e^\mathrm{SK}\geq 5\mathrm{\ MeV}-m_e)\ ,             \label{e2new}\\
\Phi_\mathrm{ES}^\mathrm{SK}    &=& 2.35 \pm 0.108\
(T_e^\mathrm{SK}\geq 8.6\mathrm{\ MeV})\ ,               \label{e3new}\\
\Phi_{B}^\mathrm{SSM}           &=& 5.93\times
 (1^{+0.14}_{-0.15})\ .                                 \label{e4new}
\end{eqnarray}

The value of $\Phi_\mathrm{CC}^\mathrm{SNO}$ in Eq.~(\ref{e1new})
is unchanged with respect to Eq.~(\ref{e1}), since no updated CC
data have been presented by the SNO Collaboration after those in
\cite{SNO1}. However, there have been several theoretical
improvements in the calculation of the $\nu_e$-$d$ CC cross
section \cite{Beac,Voge,Naka}, which can induce a few\% decrease
in the experimentally inferred value of
$\Phi_\mathrm{CC}^\mathrm{SNO}$. Pending an implementation of the
new cross sections in the official SNO data analysis, we will just
comment on their implications at the end of this section.

The value of $\Phi_\mathrm{ES}^\mathrm{SK}$ above the 5 MeV
threshold [Eq.~(\ref{e2new})] has been presented by the SK
collaboration in recent winter conferences \cite{Noon,Kosh}, and
refers to a SK detector exposure of 1496 days. The value of
$\Phi_\mathrm{ES}^\mathrm{SK}$ in Eq.~(\ref{e3new}) represents
instead the preliminary evaluation of the SK flux above the
equalized 8.6 MeV threshold \cite{Priv}, for the same exposure.
Notice that there is now a realistic $\sim 30\%$ increase of the
uncertainty when passing from Eq.~(\ref{e2new}) to
Eq.~(\ref{e3new}), to be contrasted with the previous naive
assumption of ``threshold-independent SK total flux error,''
implicit in Eqs.~(\ref{e2}) and ($\ref{e3}$).%
\footnote{The equality of the central values in Eqs.~(\ref{e2new})
and (\ref{e3new}) is instead accidental. In general, such values
can be slightly different [as it happens in Eqs.~(\ref{e2}) and
(\ref{e3})], as a result of a possible energy dependence of
$P_{ee}$ and of the experimental uncertainties associated to the
part of SK energy spectrum between the two thresholds considered.}
Therefore, the estimated SK uncertainty in Eq.~(\ref{e3new}),
although preliminary \cite{Priv}, represents an important step
forward for a more accurate model-independent analysis.

The central value and $\pm 1\sigma$ errors for
$\Phi_B^\mathrm{SSM}$ in Eq.~(\ref{e4new}) \cite{Robu} replace the
previous BP00 ones in Eq.~(\ref{e4}) \cite{BP00} as a result of a
recent, very accurate measurement of the relevant astrophysical
$S$-factor ($S_{17}(0)=22.3\pm 0.9$ eV~b \cite{S170}). By
conventionally using the central value in Eq.~(\ref{e4new}) for
normalization, we get from Eqs.~(\ref{e1new}) and (\ref{e3new})
the updated inputs for our model-independent analysis,
\begin{eqnarray}
\Phi_\mathrm{CC}^\mathrm{SNO}/\Phi^\mathrm{SSM}_B
&=& 0.295 \pm 0.025 \label{snnew}\ ,\\
\Phi_\mathrm{ES}^\mathrm{SK}/\Phi^\mathrm{SSM}_B &=& 0.396 \pm
0.018 \label{sknew}\ .
\end{eqnarray}

The difference between the above two fluxes amounts now to a
$3.3\sigma$ evidence in favor of active neutrino transitions,
which is thus slightly stronger than the one found previously
($3.1\sigma$ \cite{SNO1}).  The current situation is illustrated
in Fig.~1 (which is analogous to Fig.~2 of \cite{Ours}). The
SK~+~SNO data are well within the region where there {\em must\/}
be transitions of $\nu_e$ to $\nu_{\mu,\tau}$, independently of
the possible presence of additional $\nu_s$. To guide the eye, the
error ellipse in Fig.~1 is shown to touch the ``no active
transitions'' diagonal line only at the $3.3\sigma$ level ($\Delta
\chi^2=3.3^2$).

Assuming no sterile neutrinos, the bounds on the two parameters
 $f_B$ [now normalized to the updated flux
in Eq.~(\ref{e4new})] and $\langle P_{ee}\rangle$ are shown in
Fig.~2 (analogous to Fig.~3 in \cite{Ours}), together with the
updated SSM prediction (horizontal band). The corresponding $\pm
3\sigma$ limits set by the SK~+~SNO data are given by
\begin{eqnarray}
f_B  &=& 0.96^{+0.54}_{-0.55} ,\label{fBnew}
\end{eqnarray}
 and
\begin{eqnarray}
\langle P_{ee}\rangle  &=& 0.31^{+0.55}_{-0.16} .\label{Peenew}
\end{eqnarray}
Therefore, the new reference SSM prediction \cite{Robu} in
Eq.~(\ref{e4new}) ($f_B=1^{+0.14}_{-0.15}$ at $1\sigma$) is
strikingly confirmed, and the best-fit $^8$B $\nu_e$ flux
suppression $(31\%)$ is now slightly more pronounced  than in the
previous analysis (where it was $34\%$ \cite{Ours}).

Finally, we discuss the effects of the recent improvements in the
theoretical calculation of the $\nu_e$-$d$ CC cross section
\cite{Beac,Voge,Naka}. It has become apparent \cite{Beac} that
radiative corrections produce a few~\% increase \cite{Beac,Voge}
of the reference cross section \cite{Xsec} used in \cite{SNO1} to
extract the CC neutrino event rate. Refinements in effective field
theory calculations \cite{Naka} add a small ($\sim 1\%$)
contribution in the same direction. Assuming a representative +3\%
shift due to all such corrections (corresponding to a $+1\sigma$
increase \cite{SNO1} of the reference CC cross section
\cite{Xsec}), the values in Eq.~(\ref{e1new}) would become 3\%
smaller. Leaving Eqs.~(\ref{e3new}) and (\ref{e4new}) unchanged,
the SK-SNO flux difference would then increase from $3.3\sigma$ to
$3.6\sigma$, and the central values of $f_B$  and $\langle
P_{ee}\rangle$ would be shifted to $1.00$ and to $0.29$,
respectively, thereby reinforcing all previous conclusions.

\section{Conclusions}

The model-independent SK~+~SNO evidence for electron neutrino
flavor transitions into active states is in good shape. Its
statistical significance, evaluated at the $3.1\sigma$ level in
\cite{Ours}, is currently confirmed at the $3.3\sigma$--$3.6
\sigma$ level, in the light of recent experimental and theoretical
improvements (independently of the SSM and of the pattern of
active or sterile $\nu$ transitions \cite{Vill}). For purely
active neutrino transitions, the SK~+~SNO data
\cite{Noon,Kosh,Priv,SNO1} are in striking agreement with the
updated standard solar model $^8$B neutrino flux \cite{Robu,S170}:
$f_B=0.96^{+0.54}_{-0.55}$ ($\pm 3\sigma$). The corresponding
$\pm3\sigma$ range for the SK-SNO energy-averaged $\nu_e$ survival
probability is found to be $\langle P_{ee}\rangle=
0.31^{+0.55}_{-0.16}$.

\begin{acknowledgments}
This work was supported in part by the Italian  {\em Istituto
Nazionale di Fisica Nucleare\/} (INFN) and {\em Ministero
dell'Istruzione, dell'Universit\`a e  della Ricerca\/} (MIUR)
under the  project ``Fisica Astroparticellare.'' We thank M.\ Smy
for very useful discussions.
\end{acknowledgments}


\begin{figure}
\includegraphics[width=16cm]{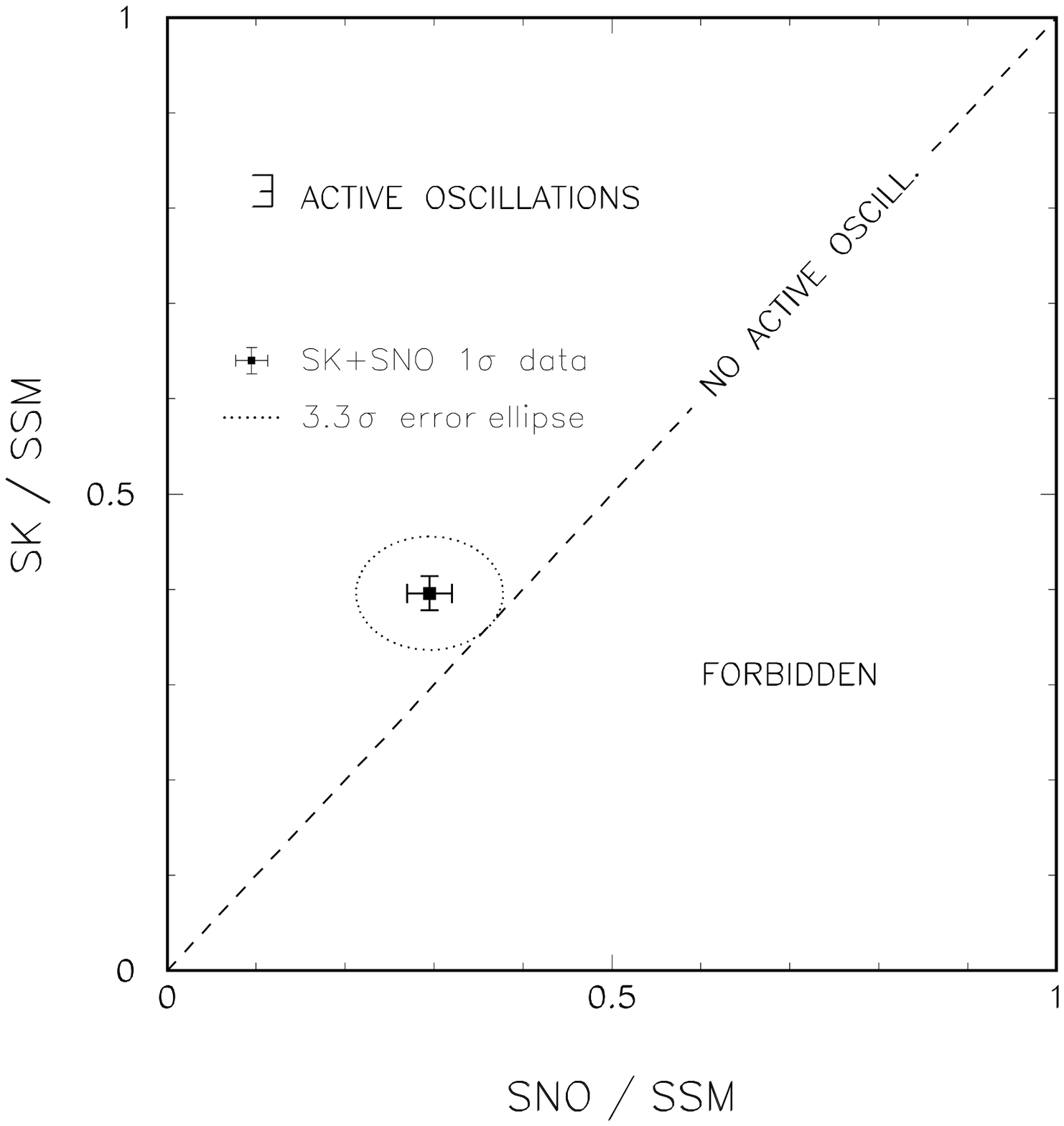}
\medskip
\caption{Current SK~+~SNO $3.3\sigma$ evidence for solar $\nu_e$
transitions into active states. Input: equalized SNO CC
\cite{SNO1} and SK ES \cite{Priv} solar neutrino event rates,
conventionally normalized to the recent SSM $^8$B flux prediction
in \cite{Robu}. An increase of the reference $\nu_e$-$d$ CC cross
section \cite{Xsec} by $\sim3\%$ (as suggested by recent
theoretical calculations \cite{Beac,Voge,Naka}) would shift the
SNO datum slightly leftwards, and would correspondingly increase
the statistical significance of the evidence to $\sim 3.6\sigma$.
See the text for details, and Fig.~2 in \cite{Ours} for a
comparison with our previous model-independent analysis.}
\end{figure}

\begin{figure}
\includegraphics[width=16cm]{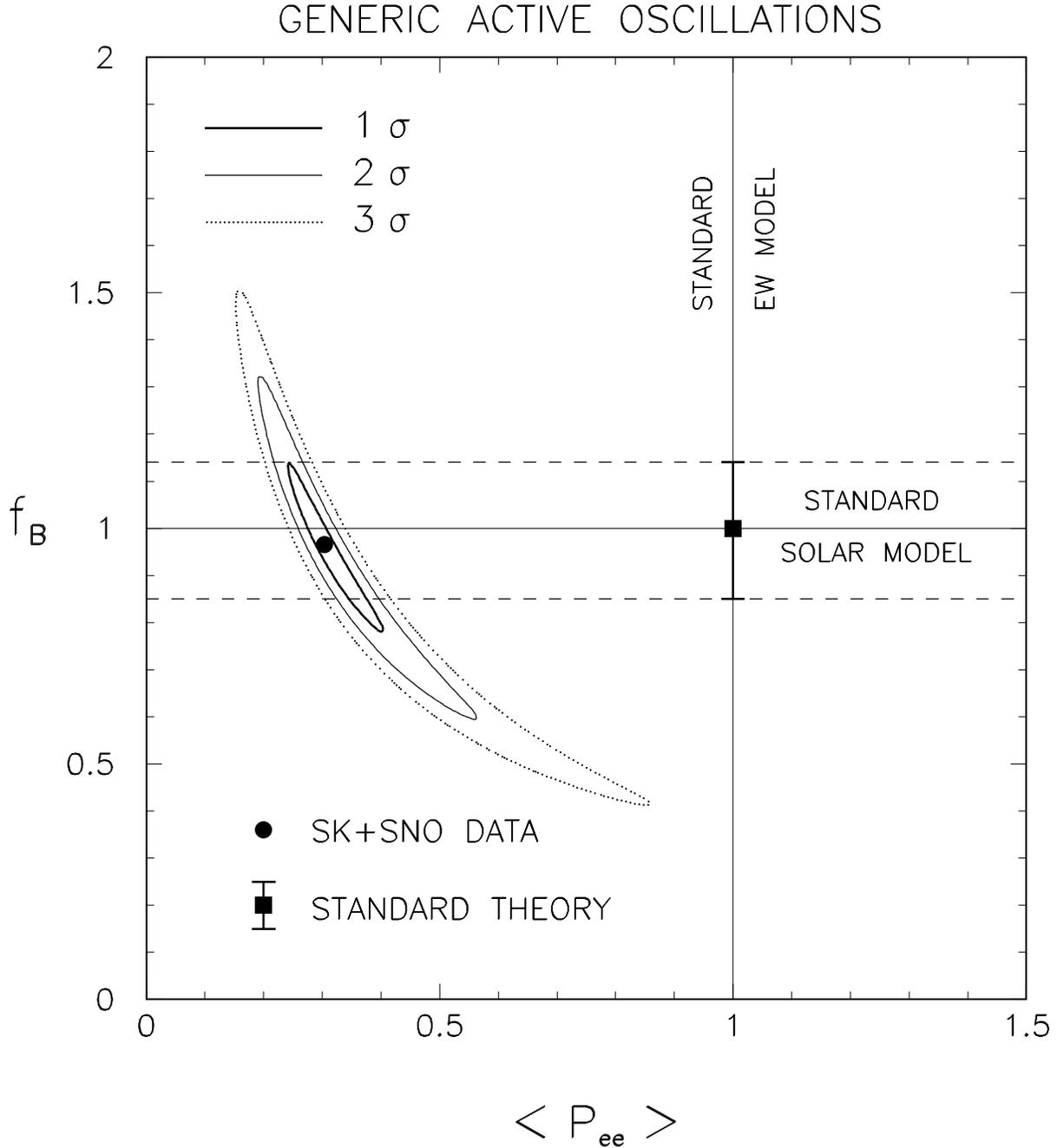}
\medskip
\caption{Current $1\sigma$, $2\sigma$, and $3\sigma$ error
contours ($\Delta \chi^2 =1$, 4, and 9), for the parameters
$\langle P_{ee}\rangle$ and $f_B$, assuming purely active $\nu$
transitions. The SSM prediction ($\pm 1\sigma$ horizontal band),
taken from \cite{Robu}, is based on the recent $S_{17}(0)$
evaluation in \cite{S170}. An  increase of the $\nu_e$-$d$ CC
cross section by $3\%$ would shift the best-fit point from
$(\langle P_{ee}\rangle,\,f_B)= (0.31,\,0.96)$ to $(0.29,\,1.00)$.
See the text for details, and Fig.~3 in \cite{Ours} for a
comparison with our previous model-independent analysis.}
\end{figure}

\end{document}